\documentclass[aps,twocolumn,epsfig]{revtex4}

\usepackage{epsfig}
\usepackage{amsmath}

\begin{document}

\title{\textbf{{\Large {Spin current rectification from a spin-biased quantum dot}}}}
\author{T. I. Ivanov}
\affiliation{Department of Physics, University of Sofia, 5 J. Bourchier Blvd., 1164 Sofia,
Bulgaria}
\received{\today}

\begin{abstract}
We compute the spin current rectification coefficient of a non-equilibrium quantum dot subject to a spin bias and an {\it ac} charge bias with small amplitude. As a function of
the position of the resonant level the spin current rectification coefficient shows a set of three peaks around the gate voltage at which the resonant or the upper level of the
dot is in the vicinity of the equilibrium Fermi level in the leads. The peak heights can be related to the average number of the quantum dot electrons. We discuss the frequency
dependence of the spin current rectification coefficient as well and emphasize the effects of the photon-assisted spin transport through the dot.

\end{abstract}
\maketitle

Recent advancement in spintronics \cite{maekawa,focus,fabian} and quantum computing \cite{awschalom,zak} results from an increasing interest in studying the spin-polarized
transport through nanostructures \cite{awschalom}. It is advantageous to manipulate spins instead of charges in materials because spins have longer coherence length as well as
relaxation time since, generally, the spin is more weakly coupled to the environment then the charge. As a part of the research effort in this field, the spin-polarized
transport through a quantum dot is extensively investigated. It has been shown that a quantum dot with Zeeman splitted states operates as a phase-coherent spin pump
\cite{mucciolo}. Spin-polarized current has been demonstrated in a quantum point contact \cite{potok1} and a Coulomb-blockaded quantum dot \cite{potok2}. The problem that
stands at the forefront in this context is to devise means of purely electrical control of the spin-polarized current. Recently, it has been proposed to couple the quantum dot
to leads which have spin-dependent electrochemical potentials, the so-called spin bias \cite{brataas}. The spin bias can be applied instead of or in addition to a charge bias
to quantum dots. It is experimentally realized as a spin accumulation at biased contacts between ferromagnetic and nonmagnetic materials or when a semiconductor lead is
illuminated by a circularly polarized light \cite{valenzuela}. The {\it dc} transport properties of spin-biased quantum dots have been investigated over the last several years
\cite{katsura}. The transport through a quantum dot subject to both constant spin bias and time-dependent charge bias has recently been addressed as well \cite{chi,ivanov10}.

Future applications of spintronic devices will require understanding of the time-dependent spin (and charge) transport through them. The work on studying the time-dependent
charge transport through nanoscopic devices (quantum dots in particular) has been an active field of research over the last 20 years \cite{haug}. Similar type of investigation
have to be performed concerning the spin transport in presence of time-dependent fields as well.

In a previous work \cite{ivanov10} we studied the linear response of a quantum dot, subject to a spin bias, to an external time-dependent charge {\it ac} bias. We examined the
behaviour of the linear response spin admittance as a function of the gate voltage and the frequency of the applied {\it ac} field. The linear response admittance is calculated
by expanding the total current through the dot to linear term in the amplitude of the {\it ac} field. In this work we extend the investigation to take into account the
non-linear terms and calculate the spin rectification coefficient of the dot. When the current-voltage characteristic of a system is non-linear or is not odd with respect to
the applied voltage a {\it dc} current can flow through the system when an {\it ac} bias acts upon it (current rectification). This effect has various applications in
electronic devices.

Spin current rectification has been studied, for example, in the case of molecular \cite{dalgleish} and quantum wires \cite{schmeltzer}. Spin-sensitive rectified current has
been experimentally observed in scanning tunneling microscope \cite{lee}. Recently, spin rectification has been studied in nonitinerant one-dimensional quantum spin chains
\cite{hoogdalem}. In this work we consider a quantum dot coupled to two leads with spin-dependent electrochemical potentials (spin bias) and additional time-dependent charge
field. We assume that the amplitude of the {\it ac} field as small and develop the tunneling current to second order with respect to the amplitude of the {\it ac} field. The
spin and charge rectification coefficients are derived from this expansion.

The Hamiltonian of the dot, coupled to two non-magnetic leads, which are considered as ideal reservoirs, with external time-dependent
charge bias applied to the dot is given by
\begin{equation}
H=H_0+ H_{ext}.
\label{eq1}
\end{equation}
$H_0$ is the Hamiltonian of the system, consisting of a quantum dot, coupled to the leads
\begin{eqnarray}
H_{0} & = &\varepsilon_c \sum \limits_{\sigma} c^+_{\sigma} c_{\sigma} +U n_{\uparrow} n_{\downarrow} + \sum \limits_{k\sigma\beta} \varepsilon_{\beta k} a^+_{\beta k\sigma}
a_{\beta k\sigma} \notag \\
& + & \sum \limits_{k\sigma\beta} (T_{\beta k} c^+_{\sigma} a_{\beta k\sigma} + H.c.),
\label{eq2}
\end{eqnarray}
where $c_{\sigma} (c^+_{\sigma})$ is the annihilation (creation) operator for an electron with spin $\sigma=\uparrow,\downarrow (+,-)$ in the dot, $\varepsilon_c$ is the
energy level and $U$ is the Coulomb repulsion energy between two electrons with opposite spins, simultaneously residing in the dot. $a_{\beta k\sigma} (a^+_{\beta k\sigma})$
is the annihilation (creation) operator for an electron in the lead $\beta=L, R$ with quasimomentum $k$ and spin $\sigma$. We assume that the electrochemical potentials in the
leads are spin dependent, that is, a spin bias is applied to the system: $\mu_{L\sigma} = \varepsilon_F + (eV_c+\sigma eV_s)/2$, $\mu_{R\sigma} = \varepsilon_F - (eV_c+\sigma
eV_s)/2$, where $V_c$ is the charge voltage, $V_s$ is the spin voltage, and $\varepsilon_F $ is the equilibrium Fermi level energy. In writing these expressions we assume that
the spin is quantized along the same axis in both leads. Finally, $T_{\beta k}$ is the tunneling matrix element. The Hamiltonian $H_{ext}$ describes the external time-dependent
perturbation. We consider the external charge {\it ac} bias to be applied to the quantum dot through the (central) gate electrode and the corresponding Hamiltonian takes the
form
\begin{equation}
H_{ext} = -e \frac {C_G} {C_{\Sigma}} u(t) \sum \limits_{\sigma} c^+_{\sigma} (t) c_{\sigma} (t).
\label{eq5}
\end{equation}
In a model of the dot with geometric capacitancies \cite{ivanov96}, $C_G, C_L, C_R$ are the capacitancies of the gate, left, and right electrodes, respectively.
$C_{\Sigma}=C_G+C_L+C_R$
is the total capacitance of the dot. The {\it ac} charge bias is $u(t)=u_0 \exp{(\rm {i} \Omega t)} + c.c.$ Note that throughout the remainder of this text (unless explicitly
written) we have omitted Planck's constant $\hbar$.

Next, we introduce the Keldysh Green's functions (GF) of the dot electrons (similar definitions hold for the lead electrons as well). There are three types of GF in the Keldysh
formulation: the retarded (advanced) and the distribution functions \cite{chow}. The retarded GF is defined as $G_{\sigma r} (t) = - {\rm i} \theta (t) \left< \{c_{\sigma} (t),
c^+_{\sigma} (0)\} \right >$ and the distribution function is $G_{\sigma <} (t_1, t_2) = {\rm i} \left< c^+_{\sigma} (t_2)c_{\sigma} (t_1) \right >$. The curly brackets denote
anticommutator and the averaging is with respect to the full time-dependent Hamiltonian of the dot. In the Keldysh technique the distribution GF depends on two times - the
"relative" time $t=t_1-t_2$ and the "center-of-mass" time $T=(t_1+t_2)/2$, the retarded (advanced) Green's functions depend only on a single time variable.

We approximate the retarded GF of the dot electrons with the simplest expression which captures the basic physics and is applicable for temperatures higher than the
characteristic Kondo temperature. The GF is obtained by applying a mean-field approximation in the equation of motion for it and the result is
\begin{equation}
G_{\sigma r} (\omega) = \frac {1-<n_{-\sigma}>} {\omega -\varepsilon_{c}+{\rm i}\gamma}+\frac {<n_{-\sigma}>} {\omega -\varepsilon_{c}-U+{\rm i}\gamma}
\label{eq8}
\end{equation}
Here, $<n_{\sigma}>$ is the average number of electrons with spin $\sigma$ in the dot and $2\gamma$ is the total elastic level width, $\gamma=\gamma_L+\gamma_R$ and
$\gamma_{\beta} = \pi \sum_k |T_{\beta k}|^2 \delta (\omega - \varepsilon_{\beta k})$. In the usually employed approximation of energy-independent tunneling matrix elements
and with a broad flat density of states for the lead electrons $\gamma_{\beta}$ are constants. Qualitatively, the spectrum of the quantum dot electrons with spin $\sigma$
consists of a lower (resonant) level with energy $\epsilon_c$ and relative weight $1-<n_{-\sigma}>$ and an upper level with energy $\epsilon_c+U$ with relative weight
$<n_{-\sigma}>$.

We calculate the distribution GF supposing that the relaxation processes in the leads are much faster as compared to those in the dot. With a spin bias applied to the quantum
dot the leads are not in equilibrium but we can assume that each spin subband has a separate equilibrium electron distribution. The leads remain in steady state in the presence
of the external {\it ac} signal because it is applied only to the quantum dot. Correspondingly, the leads' GF are given by the expressions for noninteracting equilibrium
electron systems \cite{landau}. The steady-state distribution GF for the dot electrons does not depend on the "center-of-mass" time $T$ and is given by $G_{\sigma <} (\omega)
= G_{\sigma r} (\omega) \Sigma_{0\sigma <} (\omega) G_{\sigma a} (\omega)$ where $\Sigma_{0\sigma <} (\omega) =\sum_{k \beta} |T_{\beta k}|^2 A_{\beta <} (k, \omega)$ and
$A_{\beta <} (k, \omega)$ is the distribution GF for the lead $\beta$. The average number of quantum dot electrons $<n_{\sigma}>$ is to be obtained by self-consistently solving
the equations
\begin{equation}
<n_{\sigma}> = - \int \limits^{\infty}_{-\infty} \frac {{\rm d}\omega} {2\pi} {\rm Im} \ G_{\sigma <} (\omega).
\label{eq13}
\end{equation}

The total current $I^{tot}_{\beta \sigma}, \beta=L, R$ which flows through the left/right barrier of the dot is given in the model of geometric capacitancies by
\cite{ivanov96,goldin}
\begin{equation}
I^{tot}_{L/R \sigma} = \tilde{I}_{L/R} + \frac {C_{R/L}+C_G} {C_{\Sigma}} I_{L/R \sigma} - \frac {C_{L/R}} {C_{\Sigma}} I_{R/L \sigma}
\label{eq14}
\end{equation}
where $\tilde{I}_{L/R}$ are the displacement currents through the parasitic capacitancies. They do not depend on the spin bias and are linear with respect to the applied
{\it ac} field and we shall not give the explicit expressions for them. The tunneling current through the barrier $\beta=L, R$ is given by
\begin{equation}
I_{\beta \sigma} (t) = -{\rm i} e \left < \sum \limits_k \left [T_{\beta k} c^+_{\sigma} (t) a_{\beta k \sigma} (t) - H. c. \right ] \hat{S}(t) \right >
\label{eq15}
\end{equation}
with $\hat{S}(t) = \hat{T}_p \exp{[-{\rm i} \int \limits_p {\rm d}t' H_{ext} (t')]}$. $\hat{T}_p$ is the closed-path time ordering operator \cite{chow} and the brackets
$\left< \right >$ denote averaging with respect to the time independent part of the dot Hamiltonian (namely, $H_{0}$). When the dot is subject to charge and/or spin bias both
the charge $I_c$ and the spin $I_s$ currents can flow. They are defined as follows $I_c=I_{\uparrow} + I_{\downarrow}$, $I_s=(I_{\uparrow} - I_{\downarrow})/e$ where
$I_{\sigma}=\sum_{\beta} I_{\beta \sigma}$.

The nonlinear response coefficient is calculated by expanding the tunneling current in powers of the external {\it ac} voltage bias $u(t)$. The term quadratic in the amplitude $u_0$
is obtained as
\begin{eqnarray}
& & I^{(2)}_{L\sigma}(t)   =  \frac{{\rm i} \alpha^2 e^3} {2 \hbar^3} \int \limits^{\infty}_{-\infty} {\rm d}t' {\rm d}t" u(t') u(t")  \times \notag  \\
 & & \left < \left [\sum\limits_k
T_{Lk\sigma}c^+_{\sigma} (t) a_{L k \sigma} (t) - H.c. \right]_+ \times \right .  \notag  \\
 & &  \left . \sum\limits_{\sigma' \gamma}  c^+_{\sigma' \gamma} (t')c_{\sigma' \gamma} (t') \eta_{\gamma}
 \sum\limits_{\sigma' \gamma'} c^+_{\sigma' \gamma'} (t")c_{\sigma' \gamma'} (t") \eta_{\gamma'} \right >
\label{eq16}
\end{eqnarray}
where the subscripts $\gamma, \gamma' = \pm$ refer to the positive (+) or negative (-) branch of the closed-time path, $\xi_+=\xi_-=\eta_+=-\eta_-=1$ \cite{chow}. The
time-variable of a physical quantity (the current) is always on the positive time branch. The coefficient $\alpha=C_G/C_{\Sigma}$. The reader is referred to \cite{chen} for an
introduction of the way similar expressions are dealt with. Next we perform Fourier transformation and obtain the tunneling rectification coefficient of the dot. It is defined
as the ratio of the zero-frequency Fourier transform of $I^{(2)}_{L\sigma}$ to the square of the amplitude of the external time-dependent field $a_{L\sigma}=I^{(2)}_{L\sigma}
(0)/u^2_0$. Similar definition in which the Fourier transform at frequency $2\Omega$ is involved gives the second-harmonic generation coefficient.

The explicit expression for the tunneling rectification coefficient can be cast into the form
\begin{eqnarray}
a_{L\sigma}  &=& -4 \pi {\rm i}\frac {\alpha^2 e^3} {\hbar^3} \gamma_L \int\limits^{\infty}_{-\infty}{\rm d}\omega \notag \\
& &\{ f_{L \sigma}(\omega) [G_{\sigma a} (\omega)
G_{\sigma a}(\omega+\Omega)G_{\sigma a} (\omega) - c.c. ] -  \notag \\
& & [G_{\sigma r}(\omega)G_{\sigma <} (\omega+\Omega)G_{\sigma a}(\omega)  \notag \\
& & + G_{\sigma <}(\omega)  [G_{\sigma r} (\omega)G_{\sigma r}(\omega+\Omega)  + c.c. ] ] + \notag \\
& &  (\Omega \to -\Omega) \}.
\end{eqnarray}
Here, $f_{L \sigma}(\omega)$ is the Fermi-Dirac distribution function for the electrons in the left lead with the appropriate spin-dependent chemical potential. The notation
$(\Omega \to -\Omega)$ means that one must add the same terms as those explicitly written with the substitution of $\Omega$ with $-\Omega$. Note that this expression is only
valid in the approximation used to write down the retarded GF for the dot electrons. The total rectification coefficient $a^{tot}_{L\sigma}$ is calculated from Eq.
(\ref{eq14}).

The numerical results are obtained with a set of values for the model parameters that are appropriate for the real systems: $\gamma_L=\gamma_R=\gamma, U=20 \gamma$,
temperature is taken to be $0.05\gamma$. Throughout the remainder of the text we concentrate on the case of pure spin bias applied to the dot - $V_c=0$. Let us only mention
that in this case the charge current rectification coefficient (measured in units of $\alpha^2 e^3/\hbar^3$) is generally two orders of magnitude smaller than the spin one (in
units of $\alpha^2 e^2/\hbar^3$). Consequently, the rectified charge current through the dot is very small.

In Fig. 1 we present the results for the dependence of the total spin rectification coefficient $a_{Ls}=(a^{tot}_{L\uparrow}-a^{tot}_{L\downarrow})/e$ on the position of the
quantum dot resonant level, that is, on the gate voltage applied to the dot. The behaviour of $a_{Ls}$ for a dot with applied spin bias is qualitatively similar to the
behaviour of the charge current rectification coefficient for a dot with applied charge {\it dc} bias \cite{kouwenhoven,ivanov96}. There are two characteristic three-peak
structures (one central and two side peaks) for positions of the resonant level energy close to $0$ and $-U$. They correspond to the cases when the resonant or the upper level
is near $\varepsilon_F$, correspondingly, the spin is transferred through the dot via the resonant or via the upper level. The two side peaks are of non-equal heights. The
upper side peak (the one at higher gate voltage) for the upper level and the lower side peak for the resonant level are of suppressed height. This behaviour is to be attributed
to the dependence of the relative weights of both peaks in the density of states on the average number of dot electrons $<n_{\sigma}>$ \cite{ivanov96}. Consider the two side
peaks corresponding to the resonant (higher) level. Each GF in the expression for the tunneling rectification coefficient $a_{L \sigma}$ contributes a factor $1-<n_{-\sigma}>
(<n_{-\sigma}>)$, hence $a_{L \sigma} \sim (1-<n_{-\sigma}>)^3$ in the vicinity of $\varepsilon_c \sim \varepsilon_F=0$. Similarly, in the vicinity of the upper level
$a_{L \sigma} \sim <n_{-\sigma}>^3$. The average number of dot electrons $<n_{\sigma}>$ diminishes with increasing $\varepsilon_c$ (there are more empty states in the leads
available for the tunneling electron). The spin current is given as a difference between the up-spin electron current (which depends on $<n_{\downarrow}>$) and the down-spin
electron current (which depends on $<n_{\uparrow}>$) and when a charge bias is not applied $<n_{\uparrow}> = <n_{\downarrow}>$. Hence, in this case the heights of the side peaks
in the spin current can directly be related to the average number of dot electrons as in the case of pure charge current \cite{ivanov96}.

In the case of charge current rectification the distance between the two side peaks is $2\Omega$ \cite{kouwenhoven,ivanov96}. In the present case, however, the
photon-assisted tunneling involves transitions from the dot states (for a given spin) to the spin-dependent electrochemical potentials $\mu_{L/R \sigma}$ which are separated by
the spin bias [recall that $eV_s= \mu_{L \uparrow}-\mu_{L \downarrow}= -(\mu_{R \uparrow}- \mu_{R \downarrow})$]. The peaks coming from the up-spin and from the down-spin
electrons are displaced by $\mp eV_s/2$. Consequently, the distance between the peaks is $\sim 2\Omega +eV_s$. Actually, the side peaks are somewhat more distantly separated.
As has been pointed out previously \cite{ivanov96} (in the case of charge transport through a quantum dot) the effect of the thermal fluctuations is to shift the positions of
the peaks to higher energies/frequencies.

Let us now discuss the frequency behaviour of the spin current rectification coefficient. The results are presented on Figs. 2 and 3. The presented data are distinguished by
the positions of the resonant level relative to the spin-dependent electrochemical potentials of the leads. When both resonant and the upper level are below $\mu_{L/R} \sigma$
(Fig. 2, lower panel) the spin current rectification coefficient is generally a decreasing function of the frequency because with increasing $\Omega$ the electrons cannot
follow the applied field and the current diminishes. However, there are two resonant-like features in the vicinity of $\Omega=|\varepsilon_F-\varepsilon_c|$ and
$\Omega=|\varepsilon_F-\varepsilon_c-U|$. These features are clear evidence for a photon-assisted spin transport through the dot - the electrons absorb photons in order to
tunnel through the dot and, as a result, a spin is transferred between the leads as a spin current. Similar behaviour is also observed in the case of the the so-called empty
orbital regime in equilibrium when the resonant level position is above all electrochemical potentials of the leads and there is resonant-like enhancement of the transport
coefficient for frequencies $\Omega=\varepsilon_c$ and $\Omega = \varepsilon_c+U$.

The upper panel of figure 2 presents the results for $a_{Ls}$ when the resonant level is below and the upper level is above $\varepsilon_F$. In this case $a_{Ls}$ changes sign
at $\Omega \sim 5\gamma$ and $\Omega \sim 15\gamma$. Let us assume that initially the resonant level is occupied by a down-spin electron. At frequency $\Omega=|\varepsilon_c|+
eV_s/2 \sim 5\gamma$ opens channel for inelastic transfer of down-spin electrons through the dot - a down-spin electron can absorb a photon with appropriate energy in order to
tunnel from the resonant level to the right lead. The transfer of the up-spin electrons is Coulomb-blockaded for these frequencies. For higher frequency $\Omega = \varepsilon_c
+U-eV_s/2$ the up-spin electrons can inelastically tunnel through the dot by absorbing a photon, hence, the sign of the
spin current changes. Analogously, if the dot is initially occupied by an up-spin electron at lower frequencies the up-spin electrons can inelastically tunnel from the dot to
the left lead and at higher frequencies the down-spin electrons can tunnel from the dot to the right lead giving the same sign of the spin current as in the previous case.

In Fig.3 we show the behavior of $a_{Ls}(\Omega)$ in the case when the resonant (or the upper) level is placed between the electrochemical potentials of the leads
$\mu_{L \downarrow}  < \varepsilon_c < \mu_{L \uparrow}$ (or $\mu_{L \downarrow}  < \varepsilon_c +U < \mu_{L \uparrow}$). As evidenced the spin current is negative and there
is resonant-like enhancement for $\Omega \sim 20 \gamma$. Consider the first case. If the resonant level is occupied by an up-spin electron then, at frequency $\Omega \sim
\varepsilon_c+U+eV_s/2$ a down-spin electron inelastically tunnels out of the dot to the right lead, hence, the down-spin current flows from the left to the right lead. If the
resonant level is occupied by a down-spin electron the up-spin current flows from the right to the left lead. Similar reasoning applies in the case when
$\mu_{L \downarrow} < \varepsilon_c +U < \mu_{L \uparrow}$. Apparently, if the sign of the spin bias is reversed the sign of the rectified spin current will also be reversed.

To the best of our knowledge, there are no experimental investigations of the quantum dot setup that we studied in this work. But the approximations we have made are the same
as those usually employed to discuss the spin and/or charge transport through a quantum dot. We have considered the lead electrons as noninteracting particles. This means that
we do not self-consistently determine the tunneling matrix elements and single-particle energies but simply take the corresponding quantities as input parameters. Also, we do
not take into account the effect of the time-dependent potential on the leads electrons, that is, we assume that the {\it ac} bias is completely screened in the leads.
Evidently, the frequency of the {\it ac} field must be smaller than the plasma frequency of the leads. Typically, the plasma frequency of the leads is of the order of tens of
GHz. Hence, our approach permits us to consider the response of the system to an {\it ac} field with frequency of the order of several THz which is experimentally realizable.

In conclusion, we have studied the nonlinear response of a non-equilibrium quantum dot, subject to a spin bias, to an external charge {\it ac} field. We computed the spin
current rectification coefficient of the dot as a function of the position of the resonant level and the frequency of the applied field. As a function of the resonant level
position it shows two characteristic sets of three peaks (one central and two side peaks) reflecting the spectrum of the dot. The heights of the side peaks are consistent with
the dependence of the relative weight of the resonant and of the upper level in the density of states on the average number of dot electrons. The frequency dependence of the
spin current rectification coefficient shows clear evidence of a photon-assisted tunneling - the spin is transferred through the dot upon electron absorbing a photon to tunnel
through the dot barriers.

\begin{figure}[h]
\vspace{0.5cm} \epsfxsize=7.0cm \hspace*{-1.5cm} \epsfbox{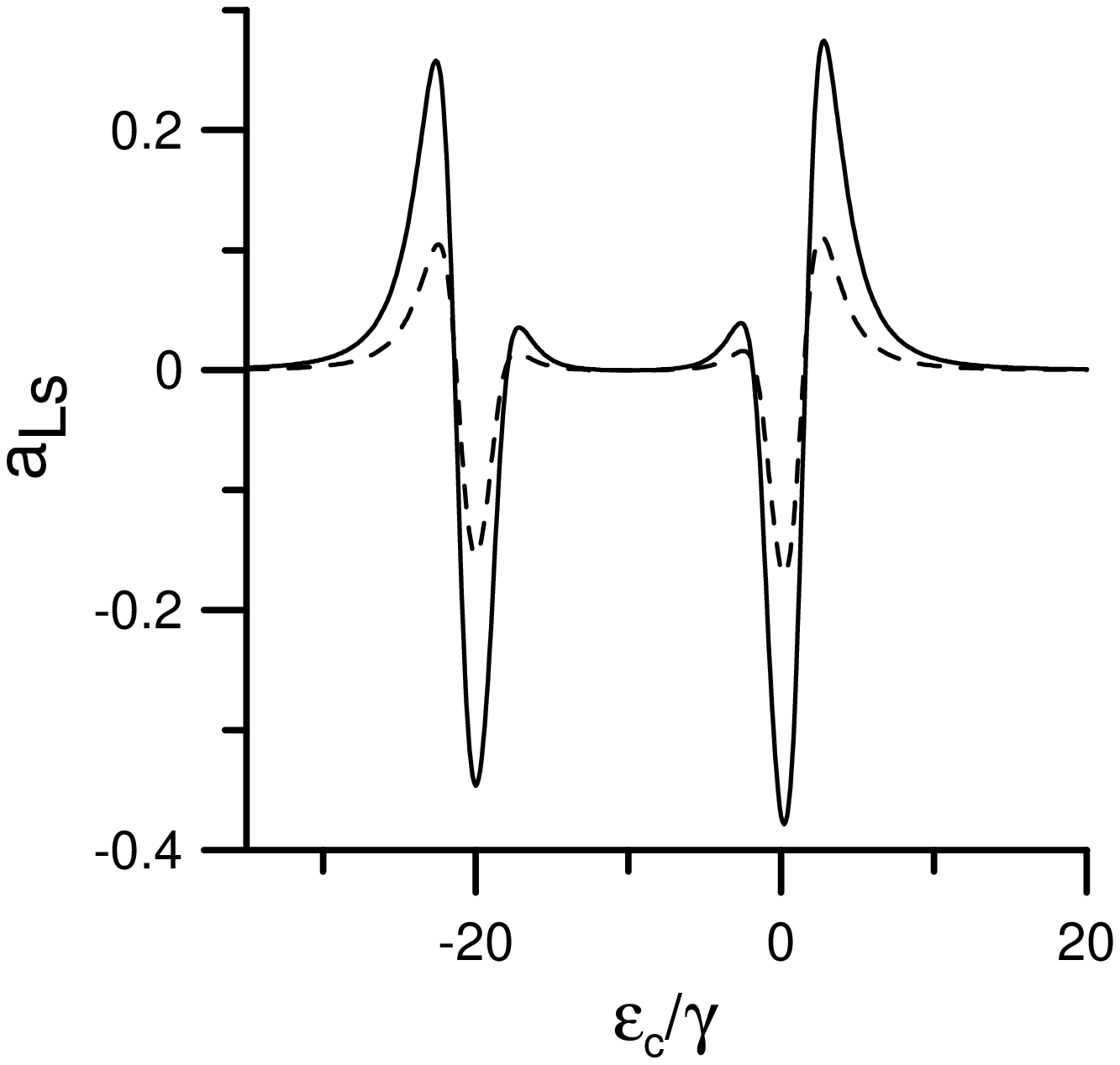}
\caption{The spin current rectification coefficient of the dot (in units of $\alpha^2 e^2/\hbar^3$) as a function of the position of the resonant
level (gate voltage) computed for frequency of the driving field $\Omega=0.6 \gamma$ and spin bias voltage $eV_s=0.5\gamma$ (solid line) and $eV_s=0.1\gamma$ (dashed line).}
\label{fig1}
\end{figure}

\begin{figure}[h]
\vspace{0.5cm} \epsfxsize=7.0cm \hspace*{-1.5cm} \epsfbox{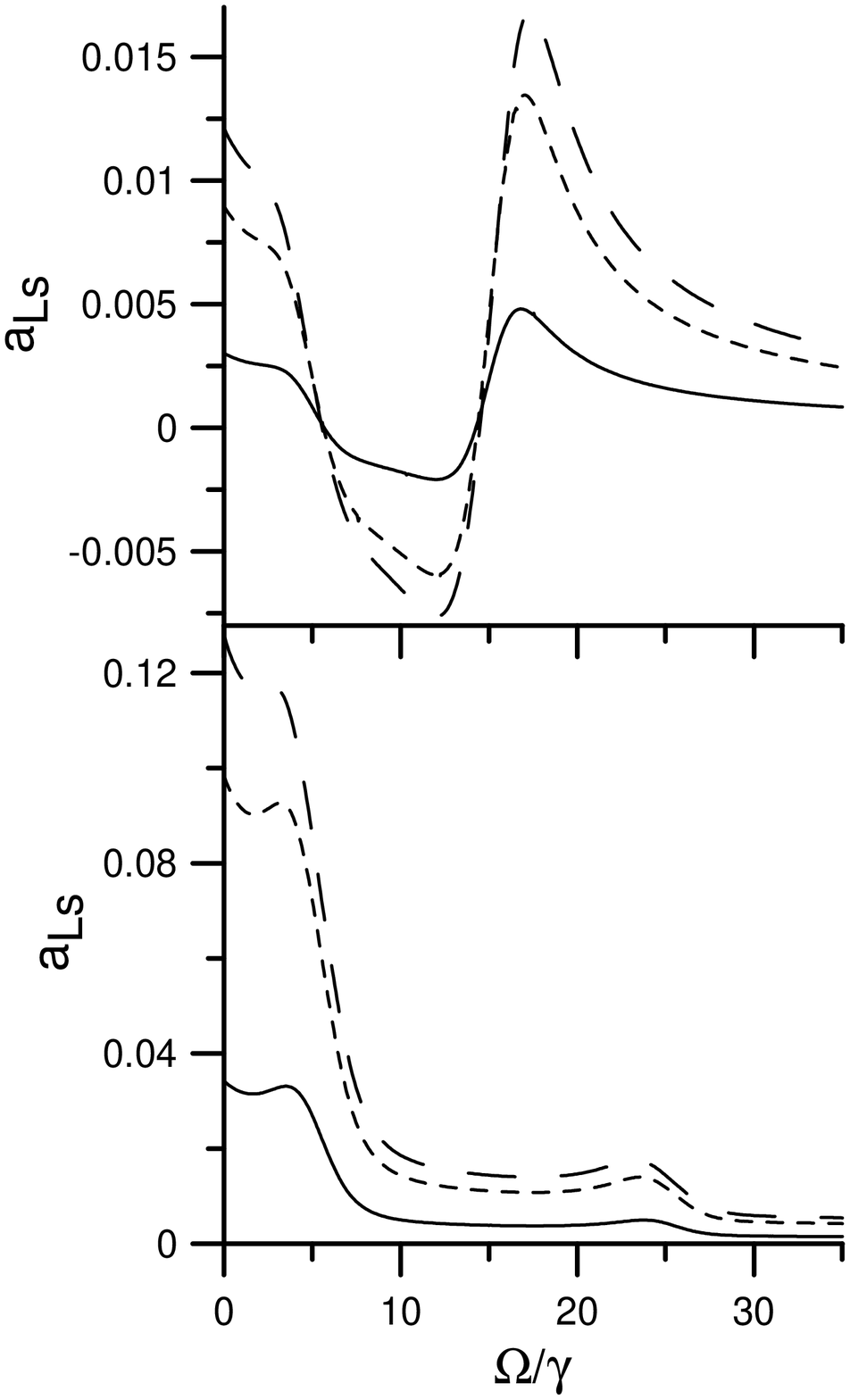}
\caption{The frequency dependence of the spin current rectification coefficient (in units of $\alpha^2 e^2/\hbar^3$) for
spin bias $eV_s=0.1\gamma$ (solid lines), $eV_s=0.6\gamma$ (dashed lines) and $eV_s=1.2\gamma$ (long-dashed lines). The resonant level energy is $\varepsilon_c=-25\gamma$
(lower panel) and $\varepsilon_c=-5\gamma$ (upper panel).}
\label{fig2}
\end{figure}

\begin{figure}[h]
\vspace{0.5cm} \epsfxsize=7.0cm \hspace*{-1.5cm} \epsfbox{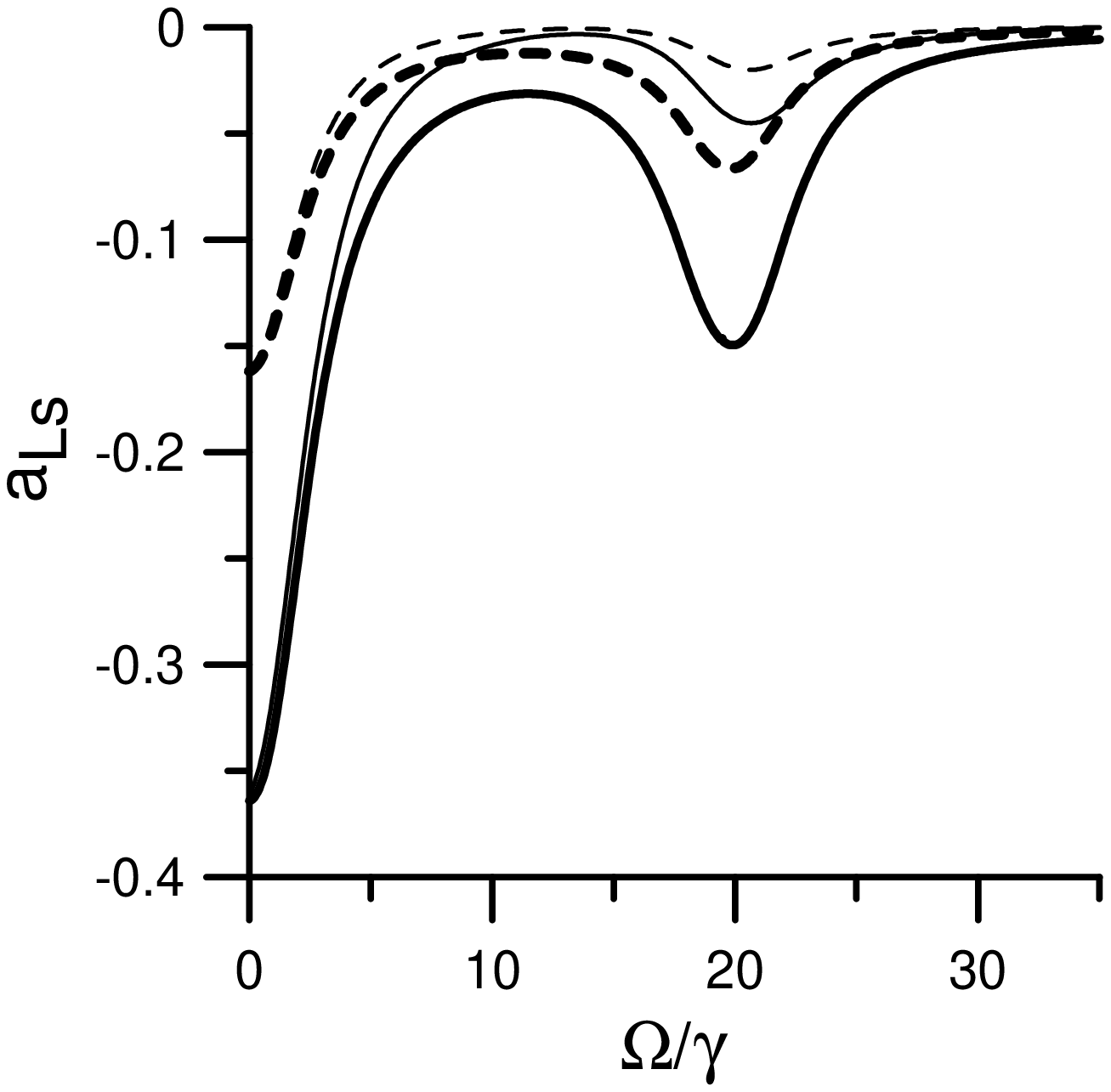}
\caption{The frequency dependence of spin current rectification coefficient (in units of $\alpha^2 e^2/\hbar^3$) for spin bias $eV_s=0.1\gamma$ (dashed lines)and
$eV_s=0.5\gamma$ (solid lines) and resonant level energy $\varepsilon_c=-0.2\gamma$ (thicker lines) and $\varepsilon_c+U=-0.2\gamma$ (thinner lines).}
\label{fig3}
\end{figure}
\end{document}